\documentstyle[prl,aps]{revtex}
\renewcommand{\baselinestretch}{1.5}
\begin{document}
\title{Quantum Zeno Dynamics and Inhibition of Geometric phase}
\author{Arun Kumar Pati and Suresh V. Lawande}
\address{Theoretical Physics Division, 5th Floor, Central Complex,}
\address{Bhabha Atomic Research Centre, Mumbai - 400 085, INDIA.}
\date{\today}
\maketitle

\begin{abstract}
Quantum  Zeno  dynamics  refers  to  the  unitary  evolution of a
quantum system interrupted by  a  sequence  of  measurements.  We
investigate  the  effect  of  sequence  of  measurements  on  the
geometric phase under quantum Zeno dynamics and show that it  can
inhibit  the  development  of  the  geometric phase under a large
number of measurement pulses. We show  that  the  path  dependent
memory of a system can be erased by a sequence of measurements, a
result  that  can  be  tested  by neutron and photon interference
experiments.

\end{abstract}

PACS  NO:           03.65.Bz\\
email:apati@apsara.barc.ernet.in

\vskip 1cm

\par
Quantum  theory  is  confronted with two types of basic dynamical
evolutions,  namely,  one  is  the  unitary, reversible evolution
leading to deterministic  solutions  and  other  is  non-unitary,
irreversible  evolution leading to probabilistic predictions. The
former is referred to as Schr\"odinger evolution and  the  latter
corresponds  to  von Neumann's collapse mechanism in the standard
`Copenhagen interpretation' of  quantum  theory.  Although  these
processes cannot be reconciled they can be allowed to compete and
when  they  do, some interesting effects occur. Unitary evolution
creates a linear superposition of all  possible  eigenstates  and
measurement  leads to the collapse of the linear superposition to
a particular eigenstate corresponding to the  eigenvalue  of  the
observable  being  measured.  On  the  other hand, if the unitary
evolution is interrupted successively at short time intervals  by
measurements  (we  call  such a dynamics as quantum Zeno dynamics
(QZD)) the quantum system continues  to  remain  in  its  initial
state.  This  inhibition  of  quantum  transition  under frequent
observations in rapid succession is called  quantum  Zeno  effect
(QZE).  Misra  and  Sudarshan \cite{ms} were the first to predict
the  QZE  for  an  unstable  quantum  system  resulting  in   the
inhibition of decay. There were various proposals \cite{hd,gm} to
test  the  prediction of QZE. Following a recent proposal of Cook
\cite{ck}, Itano et al \cite{it} have verified the inhibition  of
a  quantum  transition  in $Be^+$ ions which are laser-cooled and
trapped in a  Penning  trap.  Such  a  system  provides  a  clean
environment to test fundamental quantum predictions.

\par
   In  quantum  theory it is known that if we prepare a system in
the state $|\Psi_i>$ and a measurement is performed to see if  it
is in a state $|\Psi_f>$ at a later time, then the probability of
transition  from $|\Psi_i>$ to $|\Psi_f>$ is given by the modulus
square of the  transition  amplitude.  The  transition  amplitude
between two non-orthogonal states is given by $<\Psi_i|\Psi_f> =
|<\Psi_i|\Psi_f>| exp(i\Phi_{if})$, which has  a  modulus  and  a
phase.  Both of them have physical meaning. However, the issue of
phase is always very subtle in the quantum world  and  in  recent
years our understanding of various phases (dynamical phases, {\it
geometric  phases},  topological  phases  etc.) has been advanced
considerably. Under QZD the transition  probability  is  affected
giving  rise  to  QZE.  It is pertinent here to ask an intriguing
question: what happens to the phase under QZD?  Specifically,  we
will be concerned with the evolution of the geometric phase under
QZD.  There  are  some  deep  reasons for this. First, in quantum
theory the physical quantity of interest should not depend  on  a
particular   choice  of  gauge.  The  transition  probablity  and
geometric phase are in fact gauge invariant under  $U(1)$  action
of the state vectors. Second, both these quantities are ray space
objects  and  hence  they  are  on  the  same footing. It is thus
natural to study the behaviour of the geometric phase and not all
other phases under QZD.

   The phase $\Phi_{if}$ is quite general. If the final state has
been  obtained according to unitary and reversible dynamics, then
the transition amplitude is $<\Psi_i|\Psi_f>  =  <\Psi_i|U(t_f  -
t_i)|\Psi_i>$.   The  unitary  evolution  in  general  introduces
non-cyclic dynamical phase and geometric phase \cite{ap}. On  the
other  hand  if  the  state  of  the  system  follows non-unitary
evolution  due  to  sequence  of  measurements  of  non-commuting
observables  (hence  the  corresponding  projectors  also  do not
commute) then the system acquires a  pure  geometric  phase.  This
type  of  geometric  phase  was first encountered by Pancharatnam
\cite{sp} but later discussed in the quantum measurement  context
by  Samuel and Bhandari \cite{sb}.
The important point here is that  if  the  successive
measurements are  carried  out  along  shortest  geodesics  the
$|\Psi_i>$ is in phase with $|a>$ and  $|a>$  is  in  phase  with
$|b>$  but  $|b>$  need  not  be  in  phase with $|\Psi_i>$. This
non-transitivity  of   phase   preserving   relation   is   often
responsible for giving rise to geometric phase \cite{ja}.

     Therefore, it is  quite  clear  that  unitary  processes
introduce  dynamical  as well as geometric phases and projections
(measurements)  introduce  only  geometric  phase.  What  is  the
situation when both the evolutions are present? In this letter we
investigate   the  behaviour  of  geometric  phase  when  unitary
evolution  is  interupted  by  successive measurements of the von
Neumann type. We predict that the geometric phase of the  quantum
system  is  also inhibited like the transition probability itself
under quantum Zeno dynamics. We may refer to this effect as  {\it
quantum  Zeno  phase effect} (QZPE). Incidentally, this issue was
examined by the present authors  \cite{pl}  earlier  based  on  a
specific  model  of continuous quantum measurements \cite{ku,on}.
The dynamical equation of \cite{pl,ku,on} was non-unitary thereby
invoking the irreversibility of the measurement process but in  a
sense the description was beyond von Neumann's (orthodox) quantum
mechanics.  Another  limitation was that one measures observables
which commute with the Hamiltonian of  the  system.  The  present
proof  does not assume this restricted class of observables. Also
our earlier prediction was difficult to implement  in  laboratory
whereas  the  present  results can be tested by means of suitable
interference experiments with  neutrons  or  photons.

The geometric phase that  we are concerned with here is     quite
general, going beyond the one that was first discovered by  Berry
\cite{be}  in  his  pioneering work on quantum adiabatic theorem.
The Berry phase basically attributes a `memory'  to  the  quantum
system because it depends purely on the geometry of the evolution
path  thereby  remembering  its history. This concept was further
generalised to no-adiabatic, cyclic evolutions  by  Aharonov  and
Anandan  \cite{aa}. Further, it was generalised to non-adiabatic,
non-cyclic  evolutions  by  Samuel  and  Bhandari  \cite{sb}.   A
kinematic  and  group  theoretic  approach to geometric phase was
provided by Mukunda and Simon \cite{ms}.  More  recently,  it  is
shown  by one of the present authors \cite{ap} that the geometric
phase also arises for non-adiabatic, non-cyclic, non-unitary  and
non-Schr\"odinger  evolutions  of quantum systems. By considering
sequence of incomplete measurements Anandan and Pines  \cite{api}
have obtained geometric phase. Bhandari \cite{rb} has argued that
in   Einstein's   gedanken   experiment   {\it   for  which  path
information}  in  a  double-slit  interference   experiment   the
geometric phase plays the role of a random phase which washes out
the   interference   pattern.  Recently,  Aharonov  {\it  et  al}
\cite{at} have shown that the back reaction  due  to  measurement
process  can  induce  a  gauge potential leading to an observable
geometric phase change. In view of all these studies it is  quite
amazing  that  a counter-intuitive effect results when a sequence
of measurements  performed  on  a  quantum  system  leads  to  an
inhibition   of  geometric  phase  in  the  limit  of  continuous
measurement.

  Let  us consider a quantum system whose state vector $|\Psi(t)>
\in  {\cal  H}$  the  Hilbert  space of the system with finite or
infinite dimension. The dynamical  evolution  is  effected  by  a
Hamiltonian $H$. The state vector of the system evolves according
to Schr\"odinger equation (in the absence of measurement process)

\begin{equation}
i \hbar {d \over dt}|\Psi(t)> = H(t) |\Psi(t)>
\end{equation}

Consider an arbitray evolution (in general a non-cyclic) of a
quantum  system  for a time $T$. It is shown by one of the present
author \cite{ap,akp} that the dynamical phase and geometric phase
are given respectively, by

\begin{eqnarray}
&&\Phi_d  =  {1  \over  \hbar} \int_0^T <\Psi(t)|H(t)|\Psi(t)>~ dt, \nonumber\\
&&\Phi_g = i\int_0^T <\chi(t)|d\chi(t)>,
\end{eqnarray}
where $|\chi(t)>$ is the reference-state of the system introduced
in  \cite{ap,akp},  given  by  $|\chi(t)>  =  {<\Psi(t)|\Psi(0)>  \over
|<\Psi(t)|\Psi(0)>|}
|\Psi(t)>$.  Here,
$i<\chi(t)|d\chi(t)>$  is  basically a connection-form whose line
integral along open path in the projective Hilbert space  of  the
quantum  system gives the desired geometric phase. This geometric
phase is gauge invariant and does  not  depend  on  the  detailed
dynamics  of  the  system.  There  could be an infinite number of
open-paths in the Hilbert space but for all of them the geometric
phase  defined  above  is  the same for a given projection of the
open-path in the projective Hilbert space of the quantum  system.
This  phase  is  in  general,  {\it non-additive in nature} and
assigns a memory to the quantum system.

Imagine  a  situation  where  we have prepared our quantum system
initially in one of the eigenstates of some observable $O$, which
we are interested in measuring. The Hermitian operator $O$ has an
eigenvalue   spectrum   $\{O_n\}$   and   a   complete   set   of
eigenfunctions  $\{|\Psi_n>\}$.  The  spectrum  is  assumed to be
discrete and non-degenerate. The observable $O$ need not  commute
with  the  Hamiltonian of the system that drives the system. Now,
we allow the system to evolve under the  Hamiltonian  $H$  during
the  time  interval  $[0,T]$  such  that one performs a series of
measurements at times $\tau,  2\tau,..  (N-1)\tau,  N\tau  =  T$.
During  the  interval  $[0,\tau]$  which is very short the system
evolves unitarily. The sequence of measurements that are  carried
out  are  idealised  to  be  discrete  and  instantaneous. We are
interested in knowing how the geometric phase is affected due  to
$N$  number  of measurements on the quantum system. The fact that
the geometric phase is non-additive in nature  implies  that  the
sum  of the geometric phases acquired by the system from time $0$
to $\tau$, $\tau$ to $2\tau$ ... $(N-1)\tau$ to  $N\tau$  is  not
the  same  as the geometric phase acquired by the system from $0$
to $N\tau = T$, unlike the dynamical phase which is additive.

Here we proceed to give a simple derivation of the  result  where
the  dependence  of  the  number  of measurement on the geometric
phase can be seen explicitly. Actually, starting from the initial
state $|\Psi_n>$ the state at time $\tau$ is given by

\begin{equation}
|\Psi(\tau)> = U(\tau)|\Psi_n> = e^{-iH \tau / \hbar}|\Psi_n>.
\end{equation}

After  performing  a  von  Neumann  measurement  at  time  $\tau$
(denoted by a projection operator $P_n  =  |\Psi_n><\Psi_n|$)  to
know whether the eigenvalue of the observable is still $O_n$, the
state       is      given      by      $|\Psi_{am1}(\tau)>      =
|\Psi_n><\Psi_n|U(\tau)|\Psi_n>$, where  $|\Psi_{am1}(\tau)>$  is
the  state  of  the  system  just after first measurement at time
$\tau$. Similarly, the state of the system  at  time  $2\tau$  is
$|\Psi(2\tau)>  =  <\Psi_n|U(\tau)|\Psi_n>~U(\tau)|\Psi_n>$. When
we perform second instantaneous measurement, the state after  the
measurement     is     given     by     $|\Psi_{am2}(\tau)>     =
<\Psi_n|U(\tau)|\Psi_n><\Psi_n|U(\tau)|\Psi_n>)|\Psi_n>$.  If  we
proceed  with  $N$  number of measurement steps, the state of the
system just after $N$th measurement is given by

\begin{equation}
|\Psi_{amN}(\tau)> = <\Psi_n|U(\tau)|\Psi_n>.....<\Psi_n|U(\tau)|\Psi_n>|\Psi_n>.
\end{equation}
This  is  the final state of the system at time $T$ under quantum
Zeno dynamics. Therefore, the total phase of the  system  can  be
obtained  by  taking  the  argument  of  the inner product of the
initial and final state, in the sense of Pancharatnam. Thus,  the
total phase is given by

\begin{equation}
\Phi_t^{(N)}(T) =  arg\biggl[<\Psi_n|U(\tau)|\Psi_n>.....<\Psi_n|U(\tau)|\Psi_n>\biggr]
\end{equation}

On the other hand the dynamical phase (which is an additive quantity)
of the system under quantum Zeno dynamics is given by

\begin{equation}
\Phi_d^{(N)}(T)  = -<\Psi_n|H|\Psi_n> {T \over \hbar},
\end{equation}
which  does not depend on the number of measurements performed on
the  quantum system under going unitary evolution. Therefore, the
dynamical phase is insensitive to the quantum Zeno dynamics.

The geometric phase $\Phi_g$ is can now be expressed as

\begin{eqnarray}
&&\Phi_g^{(N)} = arg\biggl[<\Psi_n|U(\tau)|\Psi_n>.....<\Psi_n|U(\tau)|\Psi_n>\biggr] \nonumber\\
&& + <\Psi_n|H|\Psi_n> {T \over \hbar}
= N \tan^{-1}\biggl[{ - <\Psi_n|H|\Psi_n> {T \over N \hbar} \over  1 - <\Psi_n|H^2|\Psi_n> {T^2 \over 2N^2 \hbar^2} }\biggr] \nonumber\\
&& +<\Psi_n|H|\Psi_n> {T \over \hbar}.
\end{eqnarray}
The  above  expression  clearly  shows  the  {\it  dependence  of
the geometric phase on the number  of  measurements}  (i.e.,  the
number  of  times  the  system's wavefunction has been collapsed)
that has been performed on the quantum system.

Now the interesting situation arises,  when  one  takes the large
$N$ limit, i.e.,  in  the  limit  of  continuous  measurement,  a
continuous  measurement  is understood as the limit of a sequence
of  discrete,  instantaneous  measurements  when  the  number  of
measurements  tend  to infinity. Then the geometric phase behaves
as $\Phi_g^{(N)}$

\begin{eqnarray}
&&lim_{N \rightarrow \infty}
N \tan^{-1}\biggl[{ - <\Psi_n|H|\Psi_n> {T \over N\hbar} \over  1 - <\Psi_n|H^2|\Psi_n> {T^2 \over 2N^2 \hbar^2} }\biggr] \nonumber\\
&&+ <\Psi_n|H|\Psi_n> {T \over \hbar} \rightarrow 0.
\end{eqnarray}
where we have replaced $\tan^{-1}x
= x$ for small $x$. This shows that continuous measurement in
quantum   Zeno   dynamics  setting  can  completely  inhibit  the
development of the geometric phase.  Since  the  geometric  phase
attributes  a  `memory'  to  the  quantum system the above result
shows that under quantum Zeno dynamics the `memory' of  a  system
can  be  erased.  This  is  the main result of this letter. It is
immaterial which observable (commuting or  non-commuting)  of the
system is being monitored repeatedly. The prediction  of  quantum
Zeno  phase  effect requires only unitary Schr\"odinger evolution
and the projection postulate. (Recently, we \cite{apl} have shown
that the QZE in fact occurs for a wide class of systems,  obeying
non-linear,  non-unitary  equations and we hope that the QZPE can
also be predicted for non-linear systems).

The   above  idea  can  be  illustrated  with  the  neutron  spin
undergoing ``free evolution'' and measurement of its spin in  its
initial  state.  Let  us  consider a source which sends a spin-up
neutron that passes  through  several  identical  magnetic  field
regions.  The magnetic field could have components along all the three
directions $x,y$ and $z$. When the spin of neutron passes through
the magnetic field, it undergoes  precession.  During  precession
the  spin  state can acquire a geometric phase in addition to the
usual dynamical phase arising from  instantaneous  rotation.  The
interaction  of the spin with the magnetic field ${\bf B}$ can be
described by a Hamiltonian $H = \mu \sigma.{\bf B}$, where  $\mu$
is  the  magnetic moment, $\sigma$ is the Pauli spin matrices. If
the  initial  state  is  prepared  in  the  state  $|\Psi(0)>   =
|\uparrow>$, then the state after a time $t$ is given by

\begin{eqnarray}
&&|\Psi(t)> = e^{-{i\mu \sigma.{\bf B}  \over  \hbar}t}|\uparrow>
            = a(t)|\uparrow> + b(t)|\downarrow>.
\end{eqnarray}
Here,  $a(t) = (\cos{\omega t \over 2} - in_z \sin{\omega t \over
2})$, $b(t) = (n_y - in_x) \sin{\omega t  \over  2}$,  $\omega  =
{2\mu  B  \over \hbar}$ and ${\bf n} = (n_x,n_y,n_z)$ is the unit
vector in the direction of the magnetic field. If  the  evolution
is  not  interrupted  by  the  measurement process the non-cyclic
geometric phase is given by

\begin{equation}
\Phi_g(t) = -tan^{-1}(n_z \tan({\omega t \over 2}))
          + n_z {\omega t \over 2}.
\end{equation}
which can be interpreted as half the solid angle subtended by the
closed curve obtained by joining the end points of the open curve
with  a  shortest geodesic. The open curve can be parametrised by
the azimuthal angle $\phi = {\omega t \over 2}$ and polar angle  $\theta  =
\cos^{-1}(n_z  )$,  on  a  sphere which is the projective Hilbert
space for the neutron.

Let us now investigate the situation when  the  neutron  spin  is
monitored $N$ number of times  during its evolution over a time
period $T$ and under going precession in several  magnetic  field
regions  each  with a length $l$ (here $l$ is very small). Let there be a device to select
and detect the spin component of the neutron as in  the  proposed
experiment   to   test   the   QZE   by   Nakazato  {\it  et  al}
\cite{pn,hn,zh}. When netron  passes  through  a  magnetic  field
region for a small time $\tau = l/v$, with $v$ being the speed of
neutron in the field region), it undergoes rotation. The state of
the  neutron  after  passing  through  sequence  of  rotation and
projection is $|\Psi_{amN}(N\tau)>  =  a(\tau)^N|\uparrow>$.  Now
the  geometric phase shift relative to the initial state during a
quantum Zeno dynamics process is given by

\begin{eqnarray}
&&\Phi_g(N) = - N tan^{-1}(n_z tan({\omega T \over 2N})) + n_z {\omega T \over 2} \nonumber\\
\end{eqnarray}
In the limit of large number of measurements the geometric phase for
neutron  spin  goes to zero, demonstrating the quantum Zeno phase
effect, in confirmity with our prediction.

Recently  geometric phase has been measured experimentaly by Wagh
{\it  et  al} \cite{wra,wr} for neutron spin under going precession.
Here, we sketch briefly  a  neutron  interference  experiment  to
observe  this new effect (QZPE). Let there be an incident neutron
beam which is polarised along up direction and the beam is  split
coherently   into   two   parts.  In  one  arm  (say  1)  of  the
interferometer let there be several regions of magnetic field and
$N$ number of detection devices. In the other arm (say 2) of  the
interferometer  let  there  be  a  magnetic  field present over a
distance $L$ such that $L = vT$. The magnetic  field  is  applied
only   along   $z$-direction   (with   a   magnitude   equal   to
$B\cos\theta$) just to compensate the dynamical  phase  shift  in
the  interference  pattern.  The  state  of  the  neutron passing
through  the  arm  1  of  the  interferometer  is   $|\psi_1>   =
K|\uparrow>   =   a(\tau)^N|\uparrow>$,  where  $K$  denotes  the
sequence of rotation and projection operation. The state  in  the
arm   2   is   $|\psi_2>   =   exp(-i\mu   B   \cos\theta\sigma_z
T/\hbar)|\uparrow>$. When the two neutron  beams  are  recombined
the intensity of the beam is given by

\begin{eqnarray}
&&I(N) \propto |||\psi_1> + |\psi_2>||^2 \nonumber\\
&&     \propto \bigg[(1+ |a(\tau)|^{2N}) + 2|a(\tau)|^N \cos(\Phi_g(N)) \bigg].
\end{eqnarray}
This shows the interference pattern is not simply of the  type $1
+ \cos\phi$. The modification in the first term  is  due  to  the
non-unitary  operation  in  the arm 1. The interference contrast
and the phase shift of geometric origin depends on the number  of
measurements  that  have  been  performed  in  the  arm  1 of the
interferometer.  This  is  an  interesting  observation  and   by
changing  the  number  of  measurements  we  can have a different
interference  pattern.  In  the  limit   of   large   number   of
measurements   $|a(\tau)|^{2N}  =  exp(-(n_x^2+n_y^2){\omega^2  T^2
\over 4N} ) \rightarrow 1$ and the geometric phase shift tends to
zero, thus giving us a full maximum in the intereference pattern.

In conclusion we  have  predicted a new effect (QZPE), which says
that  under  frequent  measurements  the  geometric  phase of any
quantum system can  be  inhibited.  This  in  turn  implies  that
repeated measurements can erase the `memory' of a quantum system.
We  have  illustrated  the  idea  for  a neutron spin under going
succesive rotations and projections and  suggested  a  method  to
observe this effect in neutron interference experiments. This can
provide  a  new  way  of  controlling  phase  shift  by  means of
measurements which could be  of  interest  in  many  branches  of
physics.

\renewcommand{\baselinestretch}{1}

\end{document}